\documentclass[aps,prb,twocolumn,groupedaddress,showpacs]{revtex4}
\usepackage{epsfig}
\begin{document}
\title{Mott and Band Insulator Transitions in the Binary Alloy Hubbard
Model}
\author{N. Paris, A. Baldwin,
and R.T. Scalettar}
\affiliation{Physics Department, University of California, 
Davis, California 95616, USA}

\begin{abstract}
We use determinant Quantum Monte Carlo simulations 
and exact diagonalization to explore insulating behavior
in the Hubbard model with a bimodal distribution of randomly positioned local 
site energies.  From the temperature dependence of the 
compressibility and conductivity, we show that gapped, incompressible
Mott insulating phases exist away from half filling when the 
variance of the local site energies is sufficiently large.
The compressible regions around this Mott phase are metallic
only if the density of sites with the corresponding energy exceeds the
percolation threshold,
but are Anderson insulators otherwise.
\end{abstract}

\pacs{
71.10.Fd, 
71.30.+h, 
75.10.Nr, 
02.70.Uu  
}
\maketitle

\section*{Introduction}

The translationally invariant Hubbard model has long been studied as a
model of itinerant magnetism and (Mott) insulating behavior.  More
recently, the possibility of unconventional ($d-$wave) superconductivity
and spontaneously occurring charge inhomogeneities (stripes and
checkerboards) has been explored, especially in the context of high
temperature superconductivity.\cite{dagotto94}  Including disorder in
the Hubbard Hamiltonian, for example in the form of a distribution of
bond or site energies, imposes charge inhomogeneity externally and
allows for the exploration of a number of other interesting phenomena
such as the formation of Anderson insulating phases, possible
transitions to metallic behavior driven by
interactions\cite{denteneer99,balzer05}, and the influence of disorder on
magnetic correlations\cite{ulmke95}.

A particularly interesting suggestion made recently
\cite{byczuk03,byczuk04} concerns the possibility of Mott insulating
phases away from half-filling in a Hubbard model corresponding to a
binary alloy, that is, for which the probability
distribution of site energies is bimodal, 
$P(\epsilon_i) = x
\delta(\epsilon_i + \Delta/2) +(1-x) \delta(\epsilon_i - \Delta/2)$.
The idea is that if $\Delta$ is sufficiently large compared to the
bandwidth $W$, the non-interacting density of states will be split and
an insulating gap will separate two density of states peaks of weight
$x$ and $1-x$.  When an on-site repulsion $U$ is turned on, these two
peaks may in turn be Hubbard split by $U$.  Thus in the limit
$\Delta>U>W$ one can have Mott insulating phases at incommensurate
densities $\rho=x$ and $\rho=1+x$ which correspond to half-filling the
two alloy subbands.  A further interesting aspect of these Mott
insulators is that they likely occur in the absence of antiferromagnetic
ordering and its associated symmetry breaking, a phenomenon which
complicates the metal-insulator transition in the translationally
invariant Hubbard model at $\rho=1$.
Possible experimental realizations of the binary alloy Hubbard
Hamiltonian in two dimensions include Co-Fe monolayers\cite{pratzer03}.
The nature of magnetism in such systems has been explored by
first principles calculations \cite{turek94}.

The previous studies of Mott transitions off half-filling within tight binding models were with
dynamical mean field theory (DMFT).  In this letter we will re-examine
the physics of the binary alloy Hubbard model using determinant Quantum
Monte Carlo (DQMC) simulations and exact diagonalization.  While these
methods are restricted to finite size lattices, they allow us to
examine some of the aspects of the effects of randomness like Anderson
localization which are not accessible with DMFT.

The specific Hamiltonian we study is:
\begin{eqnarray}
  \label{Hamiltonian}
\hat\mathcal H=&-&t\sum_{\langle lj\rangle \sigma}
       (c_{j\sigma}^\dagger c_{l \sigma} + c_{l\sigma}^\dagger c_{j \sigma})
+U \sum_{l}  n_{l\uparrow}  n_{l\downarrow} 
\nonumber \\
&+& \sum_{l} (\epsilon_l - \mu) (n_{l \uparrow} + n_{l \downarrow})
\,\,\,.
\nonumber
\end{eqnarray}
Here $c_{l \sigma}^{\dagger} (c_{l \sigma})$ are the usual fermion
creation (destruction) operators for spin $\sigma$ on site $l$,
$n_{l \sigma} = c_{l \sigma}^{\dagger} c_{l \sigma}$ is the number
operator, and $\langle lj \rangle$ refers to near neighbor pairs on 
a two dimensional square lattice.  $t$, $\mu$ and $U$ are the electron hopping, chemical
potential, and on-site interaction strength, respectively, and
$\epsilon_{l}$ is a local site energy given by the bimodal distribution
described previously.  The bandwidth is $W=8t$ when $\Delta = U = 0$.

This paper is organized as follows:  We will first describe some of the
details of our computational methodology.  We then show results for the
density as a function of chemical potential which illustrate the
appearance of Mott plateaus off half filling, and also demonstrate the
consistency of DQMC and direct diagonalization.  Results for the participation
ratio in the non-interacting limit suggest that the Mott plateaus at
$\rho=x$ and $\rho=1+x$ could in fact be rather different, a conclusion
which we then confirm by calculating the temperature dependence of the
conductivity.  Finally, we examine the
critical hopping $t_c$ required to destroy the Mott plateau.  We
conclude by constructing the corresponding phase diagram.

\section*{Computational Methods}

We study the alloy Hubbard Hamiltonian with exact diagonalization and
DQMC \cite{blankenbecler81}.  The former approach is a standard
application of the Lanczos algorithm to determine exactly the ground
state wave function.  We use $N=8$ site lattices.  In this case $N$ is
sufficiently small that we can sum over all disorder realizations.  In
order to reduce finite size effects, we employ the boundary condition
averaging method \cite{gammel93}.  In the noninteracting limit,
averaging over different hopping phases at the boundary for a finite
lattice reproduces the thermodynamic limit spectrum exactly.  For $U$
nonzero, the finite size effects, while not eliminated, are dramatically
reduced.  Specifically, we implement a 2x4 cluster with two boundary phases 
($\phi_x$ and $\phi_y$), one boundary phase for each of the two independent, 
orthogonal boundaries.  An average over the phase space area encompassed 
by $\phi_x=\{0,2\pi\}$ and $\phi_y=\{0,2\pi\}$ was done by selecting 
10 to 100 ($\phi_x,\phi_y$) pairs.

In the DQMC approach, the partition function $Z$ is expressed as a path
integral by discretizing the inverse temperature $\beta$. The on-site
interaction is then replaced by a sum over a discrete
Hubbard-Stratonovich field \cite{hirsch85}.  The resulting quadratic
form in the fermion operators can be integrated out analytically,
leaving an expression for $Z$ in terms of a sum over all values of the
Hubbard-Stratonovich field with a summand (Boltzmann weight) which is
the product of the determinants of two matrices (one for spin up and one
for spin down).  The sum is sampled stochastically using the Metropolis
algorithm.  We present results for $6 \times 6$ lattices.  We
average over $5-10$ realizations of the local site energies.

Equal time operators such as the density and energy are measured by
accumulating appropriate elements, and products of
elements, of the inverse of the matrix whose determinant gives the
Boltzmann weight.  For the conductivity, $\sigma_{\rm dc}$, we employ an
approximate procedure\cite{trivedi96} which allows $\sigma_{\rm dc}$ to
be computed from the wavevector ${\bf q}$- and imaginary time
$\tau$-dependent current-current correlation function $\Lambda_{xx}
({\bf q},\tau)$ without the necessity of performing an analytic
continuation \cite{scalapino93},
\begin{eqnarray}
 \sigma_{\rm dc} = 
   \frac{\beta^2}{\pi} \Lambda_{xx} ({\bf q}=0,\tau=\beta/2) ~.
\nonumber
 \label{eq:condform}
\end{eqnarray}
Here $\beta = 1/T$, $\Lambda_{xx} ({\bf q},\tau) = \langle j_x ({\bf
q},\tau) \, j_x (-{\bf q}, 0) \rangle$, and $j_x ({\bf q},\tau)$ the
${\bf q},\tau$-dependent current in the $x$-direction, is the Fourier
transform of,
\begin{eqnarray}
j_x ({\bf \ell ,\tau}) =  i \sum_\sigma \, t_{{\bf \ell} + \hat{x},{\bf \ell}} \,\,
e^{H\tau}
(c^{\dagger}_{{\bf \ell} + \hat{x},\sigma}
c^{\phantom \dagger}_{{\bf \ell}\sigma}
- c^{\dagger}_{{\bf \ell}\sigma}
c^{\phantom \dagger}_{{\bf \ell}+\hat{x},\sigma}) 
e^{-H\tau}~.
\nonumber
\end{eqnarray}
This approach has been extensively tested for the
superconducting-insulator transition in the attractive Hubbard model
\cite{trivedi96}, as well as for metal-insulator transitions in the
repulsive model \cite{denteneer99,denteneer01}.

In order to get further insight into the physics of Anderson
localization, we also diagonalize the noninteracting system on lattices
as large as $N=64 \times 64$.  We characterize the properties of the
noninteracting eigenfunctions $|\phi_n\rangle$ through the scaling of
the participation ratio,
\begin{eqnarray}
{\cal P}_n = \Big( \,\,  \sum_{i=1}^{N}  \langle \, i\, | \phi_n \rangle
|^{4} \,\,  \Big)^{-1} \,\,\,.
\nonumber
\end{eqnarray}
For an eigenfunction perfectly localized at a site $i_0$, $\langle \, i
\,  | \phi_n \rangle = \delta_{i,i_0}$ we have, ${\cal P}_n=1$, while
for a perfectly delocalized eigenfunction $\langle \, i \,  | \phi_n
\rangle = 1/\sqrt{N}$, we have ${\cal P}_n=N$.  In general, ${\cal P}_n$
is a measure of the extent of the eigenfunction, that is, the number of
sites for which $\langle \, i \,  | \phi_n \rangle $ is non-negligible.

\section*{Results}

We first demonstrate the existence of the Mott and band insulating
phases by looking at the density as a function of chemical potential.
Fig.~1 shows the case $U=2$, $\Delta=4$ and $x=0.25$.  We see
a Mott insulating plateau extending from $\mu=(-\Delta-U)/2=-3$ to $\mu =
(-\Delta+U)/2=-1$, in which the total density is $\rho=x=0.25$.  At $\mu =
(-\Delta+U)/2=-1$, the lower alloy subband becomes doubly occupied and
$\rho=2x=0.5$.  A second plateau then reflects the `band' gap which must
be surpassed to begin occupying the upper alloy subband, which is, like
the lower subband, also Hubbard split.

\begin{figure}[t]
\centerline{\epsfig{figure=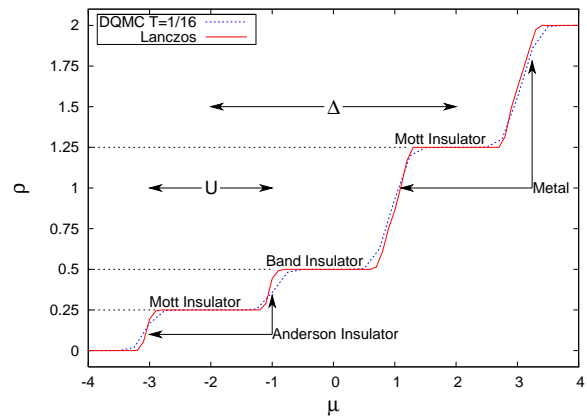,angle=-90,width=8cm}}
  \caption{(Color online)Density as a function of chemical potential for $W=0.8$,
$U=2$, $\Delta=4$ and $x=0.25$.  Solid line: ground state 
exact diagonalization of an
eight site cluster; and dashed line:  DQMC on $6\times 6$ clusters with
temperature $T=1/16$.  In the former case, results are averaged over all
configurations with two sites with $\epsilon_A=-2$ and six sites with
$\epsilon_B=+2$, and over different choices of the boundary phases.   In
the latter case, results are given for a single realization.  The two
methods give very similar results, with the DQMC somewhat rounded by
finite temperature.}
\end{figure}

In Fig.~1, the Mott plateaus revealed in $\rho\,(\mu)$ at $\rho=x=0.25$
and $\rho=1+x=1.25$ appear rather similar.  We now argue that the nature
of the states is, instead, quite different. For a square lattice in $d=2$ 
the percolation threshold is $x_c=0.5928$ \cite{percolation}.  
We therefore might expect that the
non-interacting states with site energy $\epsilon_l = -\Delta/2 = -2$,
out of which the $\rho=x$ plateau is built, are localized, since the
constitute only a fraction $x=0.25 < x_c$ of the sites in the lattice.
Meanwhile, the non-interacting states with site energy $\epsilon_l =
\Delta/2 = +2$, out of which the $\rho=1+x$ plateau is built, are
delocalized.  This is illustrated in Fig.~2 where we show the
participation ratio of the noninteracting system.  States with energies
corresponding to the upper alloy subband, which has a density 
exceeding the percolation threshold, extend over a macroscopic portion
of the lattice.  States in the lower alloy subband are localized.

\begin{figure}[t]
\centerline{\epsfig{figure=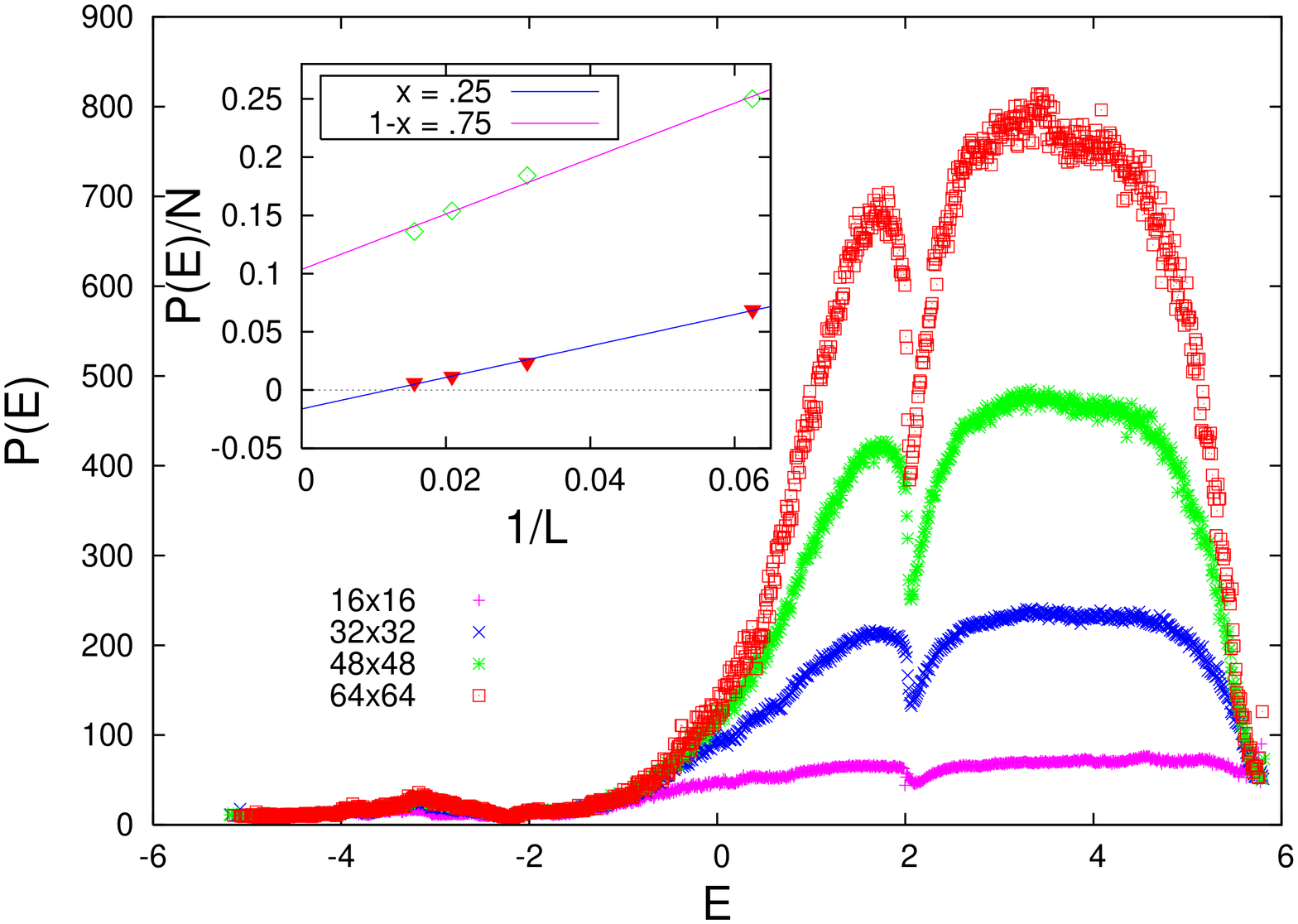,angle=-90,width=8cm}}
\centerline{\epsfig{figure=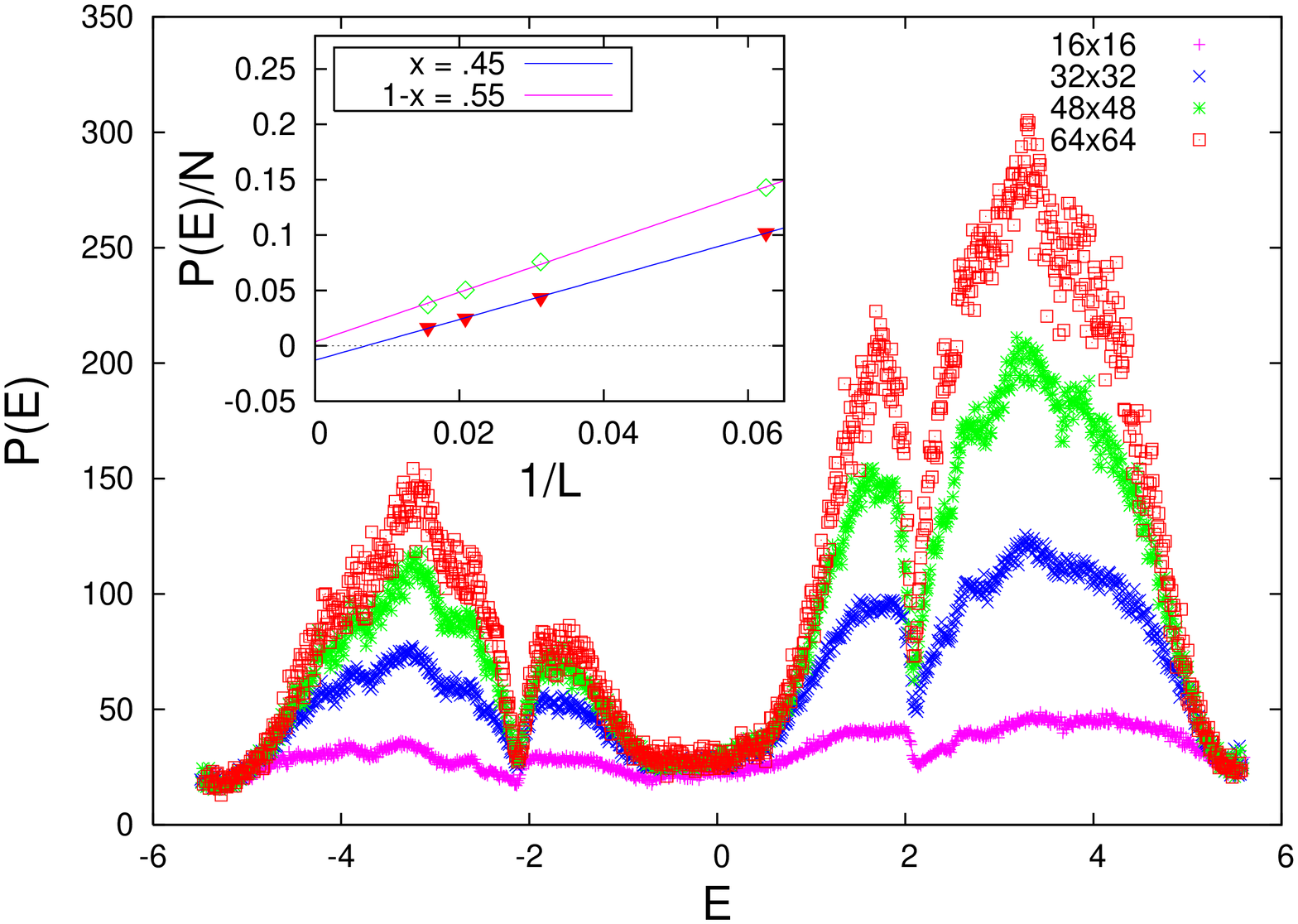,angle=-90,width=8cm}}
  \caption{(Color online)Upper panel:  Participation ratio $P$ as a function of the 
eigenenergy $E$ for the alloy Hubbard
Hamiltonian with $U=0$, $\Delta=4$, and $t=1$.  Results are shown for
lattice sizes varying from $N=16 \times 16$ to
$N= 64 \times 64$.  $P(E)$ is small
for the lower alloy band which has only $x=0.25 < x_c$ of the sites, but
$P(E)$ in the upper alloy band has a significant fraction of $N$.  The inset shows
that $P(E)/N$ scales to a non-zero value as $N \rightarrow \infty$ for the upper band
and zero for the lower band.
Lower panel:  The same for $x=0.45$.  Here both alloy subbands have a
density below the percolation threshold, and both participation ratios
scale to zero.
 }
\end{figure}

We now use the temperature dependence of the conductivity to argue that
the distinction between the two Mott insulating plateaus is preserved
when the interaction $U$ is turned on.  Fig.~3 gives $\sigma_{\rm dc}$
as a function of $\mu$ for three different temperatures $T=1/8$,
$T=1/12$ and $T=1/16$.  The conductivity is zero in both the Mott and band
insulator phases.  In the regions bracketing the upper alloy band, 
$\sigma_{\rm dc}$ is relatively large, and increases as $T$ is lowered.
That is, these regions are metallic.  For densities bracketing the lower
alloy subband, $\sigma_{\rm dc}$ is a factor of four smaller, and
increases much less noticeably as $T$ is lowered.  Because of the sign
problem. we are not able to obtain data for lower $T$.  However, we
believe that, as in the case of the Hubbard model with random site or
bond energies \cite{denteneer99,denteneer01}, the conductivity will turn
over and decrease as $T$ is lowered further, reflecting the insulating
character of the states.

\begin{figure}[t]
 \centerline{\epsfig{figure=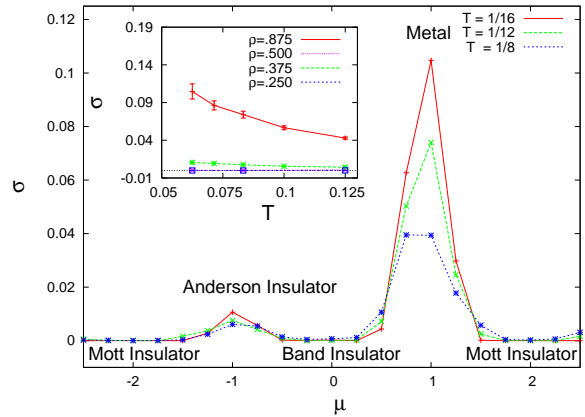,angle=-90,width=8cm}}
  \caption{(Color online)The conductivity $\sigma_{\rm dc}$ is shown as a function of
chemical potential for several temperatures.  Parameter values are as in
Fig.~1.  The upper alloy band,
whose sites have a density larger than the percolation threshold, has a
large $\sigma_{\rm dc}$, which also increases as $T$ is reduced (inset).
The insulating phases have $\sigma_{\rm dc} \approx 0$.  The
conductivity near the lower alloy subband is much smaller, and
much more weakly temperature dependent than the upper.}
\end{figure}

\begin{figure}[t]
\vskip0.2in
\centerline{\epsfig{figure=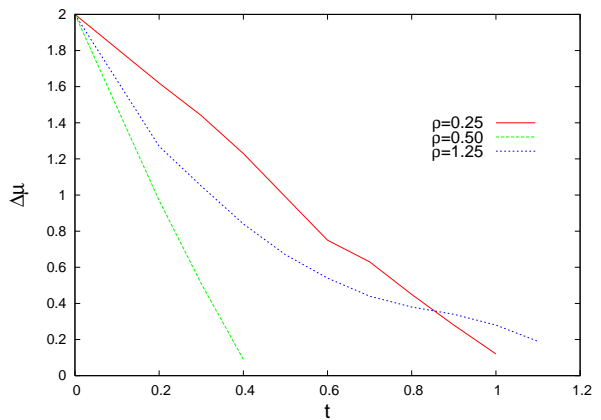,angle=-90,width=8cm}}
  \caption{(Color online)The length of the insulating plateaus $\Delta \mu$ is shown
as a function of hopping $t$ for $U=2$, $\Delta=4$ and $x=0.25$, using
exact diagonalization of $N=8$ site clusters using boundary condition
averaging.  The Mott insulator 
plateaus at $\rho=x$ and $\rho=1+x$ are more robust
than the band insulator plateau at $\rho=2x$.}
\end{figure}

We have presented DQMC results at $U=2$, $\Delta=4$ which correspond to on-site
interaction and alloy site energy separation about twice and four times
the bandwidth, respectively.  The reason for these strong coupling values
is that for larger $t$ it is not possible to reach low enough
temperatures to see clear plateaus in the density versus chemical
potential plots.  However, we argued that diagonalization and
DQMC gave consistent results (see Fig.~1), and we will now use the
former approach to generate the ground state phase diagram at weaker couplings.  As before,
we average over different boundary condition phases to reduce finite
size effects.

\begin{figure}[t]
\centerline{\epsfig{figure=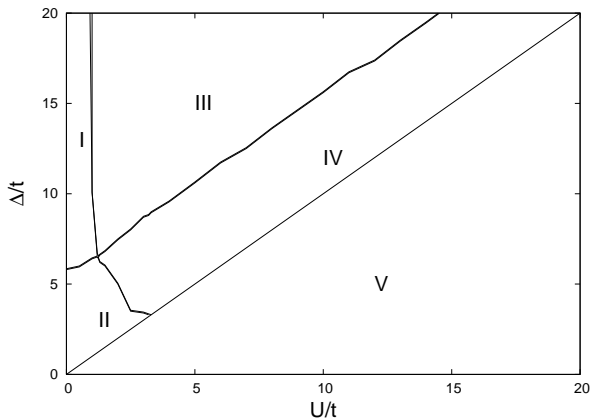,angle=-90,width=8cm}}
  \caption{Phase diagram of $U/t$ vs. $\Delta/t$.
I: Band Insulator at $\rho=0.50$ and no Mott Insulator.  II: Metallic phase.
III: Band Insulator at $\rho=0.50$ and Mott Insulators at $\rho= 0.25$ and $1.25$.
IV: Mott Insulator at $\rho=0.25$ and $1.25$ and no Band Insulator. 
V: $U>\Delta$, band insulator at $\rho=0.25$ and $1.25$ and 
Mott insulator at $\rho=1.00$.}
\end{figure}

Fig.~4 shows the length of the three plateaus as the hopping $t$ is
increased for $U=2$, $\Delta=4$, and $x=0.25$.  The two Mott plateaus
appear to vanish at roughly the same hopping strength, $t \approx 1$.
The band insulator is less robust, and is destroyed when quantum
fluctuations are only about half as strong.  The reason is that when
hopping off of one of the lower energy alloy sites for a Mott insulator you have
to pay a cost of $\Delta=4$.   However, when you hop off of one of 
these sites for  the band insulator you only have to pay the cost of
$\Delta - U=2$.  This greater ease of such charge fluctuations for the
band insulator makes its destruction by increasing $t$ occur earlier.

Similar data for other choices of $U$ and $\Delta$ allow us  to
generate the ground state phase diagram.  In 
Fig.~5 we show four possible phases for the region of which $U<\Delta$.
As in Fig.~4, large $t$ that correspond to  small $U/t$ and $\Delta/t$ 
will suppress the insulating plateaus which results in a metallic phase.
Taking the limiting case for small $U/t$ and large $\Delta/t$,we find 
one band insulating plateau at $\rho=2x$.  Inversely, for small $\Delta/t$ 
and large $U/t$, we find two Mott phases at $\rho=x$ and $\rho=1+x$.
For large $U/t$ and $\Delta/t$, both Band and Mott insulator phases
coexist.  The case for $U>\Delta$ will correspond to two band insulators
at $x$ and $1+x$ and a Mott insulator at half-filling for sufficiently large values
of $U/t$ and $\Delta/t$.

\section*{Conclusions}

In this paper we have presented DQMC and diagonalization results for the
phase diagram of the Hubbard model with binary alloy disorder.  In
agreement with previous treatments\cite{byczuk03,byczuk04}, we find Mott insulating behavior
away from half-filling when the separation of the two site energies
exceeds $U$.  We extended the earlier results to characterize the nature
of the compressible states above and below the Mott insulating plateaus
by showing that their conductivity markedly differs.  Together with the
results for the participation ratio, our data suggest that for $x<x_c$ the lowest
Mott plateau separates two compressible Anderson insulating regions,
while the upper Mott plateau separates two compressible metallic
phases.  

Our results have focused primarily on $x=0.25$, but different behavior would emerge for other values of $x$\cite{alvermann05}.  
For example, choosing a value of $x=0.70 > x_c$ would not 
only produce Mott plateaus at $\rho=0.70$ and  1.70, but also create
a lower Mott gap that are surrounded by metallic phases.  Consequently,
the upper Mott gap ($1-x = 0.30 < x_c$) would be in between two 
Anderson insulating states.

A closer examination of magnetic correlations in this model is of
interest, and will be the subject of future work.  While we expect that
disorder will suppress magnetism, as will the fact that the Mott phases
are away from commensurate fillings, it is also the case that disorder
can increase the exchange constant $J$ and hence the Ne\'el temperature
in certain circumstances \cite{ulmke95,byczuk03,byczuk04p}. Examining the real space
magnetic correlations using DQMC would be a useful complement to previous DMFT studies.

We acknowledge support from the National Science Foundation under
award NSF DMR 0312261, and useful input from M. Lovers.

\end{document}